\documentclass[a4paper,11pt,oneside]{article}

\usepackage[top=1in,bottom=1in]{geometry}
\usepackage[T1]{fontenc}
\usepackage{parskip}
\usepackage{latexsym,amsmath,amssymb}
\usepackage{times}
\usepackage{graphicx}
\usepackage{float}
\usepackage{graphicx} 
\usepackage{hyperref} 
\usepackage{amsmath}
\usepackage{authblk}
\usepackage{amssymb}
\usepackage{amsfonts}
\usepackage{makecell}    
\usepackage{blkarray}
\usepackage{booktabs}    
\usepackage{siunitx}     
\usepackage{amssymb}
\usepackage{csquotes}
\usepackage{array}
\usepackage{gensymb}
\usepackage{bbm}
\usepackage{algorithm,algpseudocode}
\usepackage{mathtools}
\usepackage{algpseudocode}
\usepackage[table]{xcolor} 
\usepackage{colortbl} 
\usepackage{geometry} 
\usepackage[backend=bibtex,
            style=numeric,
            firstinits=true,
            maxnames=6,
            sortcites=true,
            sorting=none,
            doi=false,
            url=false,
            natbib=true,
            eprint=false,
            isbn=false]{biblatex}
\usepackage{setspace}
\title{Neural Posterior Estimation on Exponential Random Graph Models: Evaluating Bias and Implementation Challenges}

\author[1]{Yefeng Fan}
\author[1,2]{Simon White}

\affil[1]{MRC Biostatistics Unit, University of Cambridge, Cambridge, UK}
\affil[2]{Department of Psychiatry, University of Cambridge, Cambridge, UK}

\begin{document}
\maketitle

\section{Abstract}
Exponential random graph models (ERGMs) are flexible probabilistic
frameworks to model statistical networks through a variety of network
summary statistics. Conventional Bayesian estimation for ERGMs
involves iteratively exchanging with an auxiliary variable due to the
intractability of ERGMs, however this approach lacks scalability to
large-scale implementations. Neural posterior estimation (NPE) is a
recent advancement in simulation-based inference, using a neural
network based density estimator to infer the posterior for models with
doubly intractable likelihoods for which simulations can be generated. While
NPE has been successfully adopted in various fields such as cosmology,
little research has investigated its use for ERGMs. Performing NPE on
ERGM not only provides a differing angle of resolving estimation for
the intractable ERGM likelihoods but also allows more efficient and
scalable inference using the amortisation properties of NPE, and
therefore, we investigate how NPE can be effectively implemented in ERGMs.

In this study, we present the first systematic implementation of NPE
for ERGMs, rigorously evaluating potential biases, interpreting the
biases magnitudes, and comparing NPE fittings against conventional
Bayesian ERGM fittings. More importantly, our work highlights
ERGM-specific areas that may impose particular challenges for the
adoption of NPE.

\textbf{Keywords:} ERGM, NPE

\newpage

\section{Introduction}

Exponential random graph models (ERGMs) provide a well-rounded
framework to infer properties of statistical networks by considering a
range of topological features through the incorporation of network
summary statistics. ERGMs have shown great advantages in being
flexible and can capture a wide range of network dependencies, and
they have been applied to various fields including sociology and neuroimaging
\citep{complexity222,book2,robins2011exponential,
  Lehmann665398}. While past ERGM research mainly perform under the
frequntist setting with well-developped package to facilitate its
wider implementations \citep{package1,package2,package3}, the
increasingly adopted Bayesian ERGMs allow fully probabilistic
evaluation of uncertainty, inclusion of prior knowledge, and remedies
to the ERGM specific degeneracy issues \citep{caimo2011bayesian,
  caimo2017bergm}. Implementation of Bayesian ERGM is subject to the
doubly intractability issue, rooted from the intractable normalising
constant in the ERGM likelihood due to the combinatorially large graph
space. Fitting of Bayesian ERGMs requires the use of the exchange
algorithm \citep{caimo2011bayesian, caimo2017bergm}, using an
auxiliary variable to explore the relation between the graph and
parameter space and iteratively consider exchanging with the estimated
parameters. It uses the augmented posterior trick to allow
cancellation of the intractable normalising constants in the ERGM
likelihood.

More generally, the exchange algorithm belongs to the likelihood-free
Bayesian inference or simulation-based inference. Approximate Bayesian
computation (ABC) is a well-known likelihood-free approach used in
past studies \citep{aos}. More recently, a class of likelihood-free
approaches called the Neural Posterior Estimation (NPE) has taken an
important role in various studies \citep{papamakarios2016fast,
  greenberg2019automatic, dax2021real,
  papamakarios2019sequential}. NPE is a neural network-based approach
to train a conditional estimator to model the Bayesian posterior
distribution \citep{papamakarios2016fast}. In comparison with
traditional ABC and the ERGM exchange algorithm, NPE can flexibly
model high-dimensional parameter space using expressive neural network
architectures such as normalising flows. In particular, NPE shows
better amortisation, referring to the little cost of re-fitting the
model upon inference on a new observation, leading to more accurate
and scalable inference \citep{pesonen2023abc, avecilla2022neural}. NPE
has since been applied to a range of studies including cosmology and
neuroscience \citep{fengler2021likelihood, vasist2023neural,
  gonccalves2020training}. Furthermore, different NPE variants have
been proposed \citep{ward2022robust}, and a well-rounded package has
been developed \citep{tejero2020sbi}. Among the NPE variants, a
particularly important extension is the sequential neural posterior
estimation (SNPE), which iteratively refines the conditional estimator
towards the target posterior \citep{greenberg2019automatic}.

In the past,  variational approximations have been performed on
ERGM. Examples are Kernel-based ABC and Copula ABC
\citep{yin2020kernel, karabatsos2024copula}. However, NPE has not been
performed on ERGM. To date the only related work is by \citet{fen2024simulation}; focusing on using SNPE
to estimate general structural network models. While insightful,
\citet{fen2024simulation} does not address ERGMs directly, concluding
that trouble arises in estimating for ERGMs (without presenting any 
fitting results or discussing specific details). Another
very recent and partially related work is performed by \citet{mele2025estimating}; focusing on the frequentist setting rather
than Bayesian. \citet{mele2025estimating} propose training a feedforward neural network to infer the
estimation. This approach functions similarly as the general NPE
literature but with a different objective function, contributing as an
alternative to the conventional Markov Chain Monte Carlo maximum
likelihood estimation (MCMCMLE) used in the frequentist ERGM
estimations \citep{ergmfit,package2,betterfit}. While important to the
development of NPE in ERGMs, their work fundamentally differs from our study in terms of both
research interest and inference paradigm

Therefore, our study contributes to fill the above-mentioned research
gap. Firstly, to rigorously and systematically assess the overall
appropriateness of adopting NPE on ERGMs; for which we adopt synthetic
data analysis to evaluate the potential biases of using NPE on
ERGMs. Secondly, our study aims to quantify the magnitudes of the
abstract biases evaluated in the parameter space through an
interpretative way. Thirdly, to assess and compare the model fitting
performance between NPE fittings and conventional Bayesian ERGM
fittings; not only the standard NPE but also the use of SNPE on
ERGMs. Finally, and most importantly, we aim to highlight
ERGM-specific aspects that are more likely to pose challenges for the
NPE and SNPE implementations. This contributes to building a solid
foundation of implementing NPE-based inference approaches on ERGMs and
provides guidelines for diagnosing the NPE fitting issues in future
implementations.

\section{Methodology}
\subsection{Exponential Random Graph Model}

ERGM characterises its probability distribution of graphs such that
the topological structure of an observed network $y$ can be explained
by a set of summary statistics or sufficient statistics
$h(y)\in \mathbb{R}^{p}$, which is flexible to include a wide range of
summary statistics and nodal covariates \citep{complexity222,book2},

\begin{equation*}
  Y \sim \text{ERGM}(\theta) \quad \text{s.t.} \quad p(Y=y|\theta)=\dfrac{\exp(\theta h(y))}{c(\theta)},
\end{equation*}
where $\theta \in \mathbb{R}^{p}$ is a vector of parameters associated with the summary statistics, and $c(\theta)$ is the normalising constant of the form,
\begin{equation*}
  c(\theta)=\sum_{y\in \mathcal{Y}} \exp(\theta h(y)).
\end{equation*}
The normalising constant is typically intractable due to the combinatorial large graph space $\mathcal{Y}$, and this makes the sampling distribution analytically intractable, which forms the main difficulty in fitting an ERGM.\\
In the Bayesian setup, a common ERGM estimation procedure is the Exchange Algorithm \citep{caimo2011bayesian, caimo2017bergm}. It iteratively explores the mapping between the parameter space and the graph space by introducing an auxiliary variable to get around the doubly intractability problem, allowing cancelling out the intractable normalising constants. More specifically, given a prior distribution $\pi(\theta)$, the algorithm instead targets an argumented posterior
$$\pi(\theta,\theta',y,y')\propto \pi(\theta'|y')\pi(\theta|y)g(\theta'|\theta),$$
where $\pi(\theta|y)$ is the original posterior. The auxiliary variable $y' \sim p(\cdot | \theta')$ is simulated with the ERGM density characterised by $\theta'$, where the simulation algorithm is in Supplementary material. We use a symmetric normal proposal $g(\theta'|\theta)=\mathcal{N}(\theta, \Sigma)$ with arbitrary proposal variance $\Sigma$. We can accept $\theta'$ with the acceptance probability
\begin{equation}
\label{eq:ar_equation}
\begin{aligned}
\text{Acceptance Ratio} 
&= \frac{\pi(\theta' \mid y)}{\pi(\theta \mid y)} \cdot \frac{\pi(y' \mid \theta)}{\pi(y' \mid \theta')}\\
&= \frac{\exp\!\bigl(\theta' h(y)\bigr)\,\pi(\theta')\,c(\theta)\,c(\theta')\exp\!\bigl(\theta h(y')\bigr)}
         {\exp\!\bigl(\theta h(y)\bigr)\,\pi(\theta)\,c(\theta')\,c(\theta)\exp\!\bigl(\theta' h(y')\bigr)}\\
&= \frac{\exp\!\bigl(\theta' h(y)\bigr)\,\pi(\theta')\exp\!\bigl(\theta h(y')\bigr)}
         {\exp\!\bigl(\theta h(y)\bigr)\,\pi(\theta)\exp\!\bigl(\theta' h(y')\bigr)}
\end{aligned}
\end{equation}
The estimation procedure is summarised in Algorithm~\ref{alg:approximate_exchange}.\\

\begin{algorithm}

\caption{Exchange Algorithm}
\label{alg:approximate_exchange}
\begin{algorithmic}[H]
\State \textbf{Initialise:} Set $T$ and initialise $\theta^{(1)}$
\For{$t=1,...,T$}
\State Propose $\theta_{j}' \sim \mathcal{N}(\theta_{j}^{(t)}, \Sigma)$
\State Simulate $y_{j}' \sim p(\cdot|y_{j-1},\theta_{j}')$ 
\State Accept the proposal $\theta_{j}^{(t+1)}=\theta_{j}'$ with Acceptance ratio in Equation~\eqref{eq:ar_equation}
\EndFor
\end{algorithmic}
\end{algorithm}

\subsubsection{Summary Statistics and Network Sizes}
In our analysis, we consider summary statistics: the number of edges, GWESP, and GWNSP, which are defined as:
$$h_{GWESP}(y) = \exp(\tau)\sum_{i=1}^{n-2}\left(1 - \left(1 - \exp(-\tau)\right)^{i}\right)p_{i},$$
$$h_{GWNSP}(y) = \exp(\tau)\sum_{i=1}^{n-2}\left(1 - \left(1 - \exp(-\tau)\right)^{i}\right)np_{i},$$
where $p_{i}$ and $np_{i}$ are the number of connected and non-connected vertex pairs sharing $i$ neighbours, respectively.

While summary statistics selection and their impact on NPE is not of research interest, we choose this combination of summary statistics based on the fact that it has been successfully adopted on different past studies and have important inference implications on network efficiencies \citep{complexx,Lehmann665398}, and we set the decay parameter $\tau$ the same as the previous studies at 0.75 \citep{Lehmann665398}.

Furthermore, we consider 90-vertex networks in this study, which aligns with the past neuro-imaging ERGM implementations on the real-world data from the Cambridge Centre for Ageing and Neuroscience project \citep{shafto2014cambridge,Lehmann665398}. Such selected network size ensures implementation practicality and adequate results generalisability to most moderate-sized real-world network applications.

\subsection{Neural Posterior Estimation}\label{c_npe_s_npe}

\subsubsection{Data Definition and Sufficiency}
Before the introduction of the NPE, we specify the definition of (observed) data for NPE. More specifically, the observed data are the summary statistics $h(y)$ instead of the network $y$ itself. Because it is challenging for neural networks to model high-dimensional outputs \citep{labbe2009learning}, and graph isomorphism (the same graph can be labelled differently) further adds to the complexity \citep{MCKAY201494, ireegular}.

We can model the relationship between model parameters and summary statistics $h(y)$ instead of modelling between parameters and realised graph $y$ using the Bayes sufficiency of $h(y)$ for parameter $\theta$ (see proof in the Supplimentary Material),
$$\pi(\theta|y)=\pi(\theta|h(y)).$$
Therefore, for inference on $\theta$, we do not need additional information from the network $y$. To differentiate, we use $x$ to denote the statistics such that $x=h(y)$, and $x_{obs}$ represents the observed network statistics.

\subsubsection{Amortised Neural Posterior Estimator}
NPE was first proposed by Papamakarios et al. and targets learning a parametric approximation to the exact posterior distribution $\pi(\theta|x)$ using a neural network \citep{papamakarios2016fast}.

In the first simulation stage, a set of $B$ data-parameter pairs $\{\theta_{b}, x_{b}\}_{b=1}^{B}$ are simulated, forming our training dataset. The parameters may be simulated from any proposal distribution $\tilde{p}(\theta)$ but often chosen as the prior (i.e., $\tilde{p}(\theta)=\pi(\theta)$),
$$\theta_{b} \sim \tilde{p}(\theta),$$
$$x_{b} \sim p(x|\theta_{b}).$$
The simulation pairs are used to train a conditional density estimator denoted as $q_{\phi}(\theta|x)$, where $\phi$ parametrises the inference neural network. In this paper, we mainly use the masked autoregressive flow (MAF) as the conditional density estimator \citep{papamakarios2017masked}, and we summarise more conditional density estimators and their usage in the Supplementary Materials as it is not of our direct research interest.

In the next training step, the conditional density estimator is trained through $\phi$ by maximising the likelihood of the data-parameter pairs. More specifically, the objective function is,
$$\mathcal{L}(\phi) = -\mathbb{E}_{\theta, x} [ \log q_\phi(\theta | x) ],$$
where the expectation is taken with respect to the $p(\theta, x)=p(x|\theta)\tilde{p}(\theta)$. The objective function can be approximated using the training data such that we minimise the negative log-likelihood,
$$\mathcal{L}(\phi) \approx -\sum_{b=1}^{B}log q_{\phi}(\theta_{b}|x_{b}).$$
Papamakarios et al. proved that as $B \rightarrow \infty$, the training density estimator becomes \citep{papamakarios2016fast},
\begin{equation}
\label{eq:c_npe_s_npe_proposal_poster}
q_{\phi}(\theta|x) \propto \dfrac{\tilde{p}(\theta)}{\pi(\theta)}\pi(\theta|x).
\end{equation}
If we simulate directly from the prior distribution (i.e., $\tilde{p}(\theta)=\pi(\theta)$), the trained density is a direct estimator of the posterior distribution. For the implementation of amortised NPE, we consider only simulating from the prior distribution (i.e., setting $\tilde{p}(\theta)=\pi(\theta)$).

In the final inference step, the trained conditional density estimator that targets the posterior distribution is conditioned on a particular observation $x_{obs}$ such that the parameter inference is performed on $q_{\phi}(\theta|x_{obs})$. The trained density estimator is amortised, meaning inference can be performed on different observations without the need to re-fit the model.\\

\subsubsection{Sequential Neural Posterior Estimation}\label{c_npe_s_snpe}

Despite the advantage of amortisation of the previously introduced amortised NPE, the trained conditional density estimator maybe subject to less accuracy if the training dataset contains insufficient data-parameter pairs within the target posterior parameter domain.

Therefore, we move to SNPE, which uses the trained density estimator as the proposal distribution for the next round of training to refine the training dataset and focus more on the target posterior domain, which, inevitably, requires adjustment of proposal distribution $\tilde{p}(\theta)$. However, from Equation~\eqref{eq:c_npe_s_npe_proposal_poster}, we can see that using a proposal distribution $\tilde{p}(\theta)$ different from the prior distribution $\pi(\theta)$ leads to the trained density estimator no longer yielding the target posterior.

More specifically, given a proposal density, the density estimator targets instead a density referred to as the proposal posterior $\tilde{p}(\theta|x)$, and this can be seen from the full expansion of Equation~\eqref{eq:c_npe_s_npe_proposal_poster},
\begin{equation}
\label{eq:c_npe_s_snpe_base_equation}
\tilde{p}(\theta|x) = \pi(\theta|x) \dfrac{\tilde{p}(\theta)p(x)}{\pi(\theta)\tilde{p}(x)},
\end{equation}
with 
$$\tilde{p}(x)=\int_{\theta}\tilde{p}(\theta)p(x|\theta).$$
SNPE proposed by Greenberg et al. uses $q_{\phi}(\theta|x)$ to target the true posterior $\pi(\theta|x)$ and considers $\tilde{q}_{\phi}(\theta|x)$ to target the proposal posterior $\tilde{p}(\theta|x)$\footnote{Note that we slightly abuse the notation of $\phi$ here. Only $\tilde{p}(\theta|x)$ is a conditional density estimator and $\tilde{q}_{\phi}(\theta|x)$ is achieved through the transformation of $\tilde{p}(\theta|x)$.} \citep{greenberg2019automatic},
\begin{equation}
\label{eq:c_npe_s_snpe_main_equation}
\tilde{q}_{\phi}(\theta|x)=q_{\phi}(\theta|x)\dfrac{\tilde{p}(\theta)}{\pi(\theta)}\dfrac{1}{Z_{\phi}(x)}.
\end{equation}
The $Z_{\phi}(x)$ is the normalising constant,
$$Z_{\phi}(x)=\int_{\theta} q_{\phi}(\theta|x)\dfrac{\tilde{p}(\theta)}{\pi(\theta)}.$$
This format aligns with Equation~\eqref{eq:c_npe_s_snpe_base_equation}, because when $q_{\phi}(\theta|x)=\pi(\theta|x)$, then $\dfrac{1}{Z_{\phi}(x)}=\dfrac{p(x)}{\tilde{p}(x)}$, re-showing Equation~\eqref{eq:c_npe_s_snpe_base_equation}. When $q_{\phi}(\theta|x)$ shall estimate the target posterior, $\tilde{q}_{\phi}(\theta|x)$ estimates for the proposal posterior, and this enables direct \enquote{read off} the correct conditional density estimator without the need for post-hoc solving. However, the integral in $Z_{\phi}(x)$ is not always evaluatable, and it depends on the choice of the proposal distribution and the type of conditional density estimator. MAF is a case where $Z_{\phi}(x)$ is not analytically evaluatable. To overcome such an issue, Greenberg et al. proposed to use an atomic loss to enable flexibility to incorporate various proposal distributions and density estimators, which includes MAF. However, such flexibility comes with the cost of potential risks of leakage issues, referring to the situation where the estimated posterior distribution produced by a neural network assigns a lot of probability mass to regions of the parameter space that should have little probability under the true posterior distribution. We include a summary of the atomic loss in the Supplementary Materials, and the limitation of the leakage issue will be discussed upon implementation.

When training the SNPE across rounds, we target the objective function,
$$L(\phi)=-\sum_{b=1}^{B} log \tilde{q}_{\phi}(\theta_{b}|x_{b}).$$
Using the proposition proved by Papamakarios et al. (Equation~\eqref{eq:c_npe_s_npe_proposal_poster}) \citep{papamakarios2016fast,greenberg2019automatic}, as $B \rightarrow \infty$, $\tilde{q}_{\phi}(\theta|x) \rightarrow \tilde{p}(\theta|x)$ and thus $q_{\phi}(\theta|x) \rightarrow p(\theta|x)$.

Now SNPE is able to \enquote{correct} the estimation during the training process such that it targets the correct posterior instead of the proposal posterior. To iteratively refine the density estimator, SNPE can aggregately consider the loss across multiple iterations. More specifically, we start with the initial round by considering the proposal distribution the same as the prior distribution. Then, we use the trained $q_{\phi}(\theta|x)$ as proposal distribution in further rounds. Notably, we re-use the data from previous iterations in each iteration. However, due to iterative refinement, the amortisation properties are lost (i.e., SNPE needs to be retrained when inferring on a different observation). We outline SNPE in Algorithm~\ref{alg:c_npe_s_snpe}.

\begin{algorithm}
\caption{Sequential Neural Posterior Estimation (SNPE)}
\label{alg:c_npe_s_snpe}
\begin{algorithmic}
\State \textbf{Initialise:} 
        Set initial proposal as $\tilde{p}_1(\theta)=\pi(\theta)$ and choose $T$\\

    \For{$t = 1$ to $T-1$}

        \State Sample $\theta_{b,t} \sim \tilde{p}_t(\theta)$ for $b = 1, \dots, B$
        \State Simulate $x_{b,t} \sim p(x|\theta_{b,t})$ for $b = 1, \dots, B$
        \State Train $\phi = \operatorname*{arg\,min}_\phi \sum_{t=1}^T \sum_{b=1}^B -\log \tilde{q}_{\phi}(\theta_{b,t}|x_{b,t})$ 
        \State Set proposal $\tilde{p}_{t+1}(\theta)=q_\phi(\theta|x_{\text{obs}})$ 
    \EndFor
    \State \textbf{Return:} $q_\phi(\theta|x_{\text{obs}})$
\end{algorithmic}
\end{algorithm}

\section{Implementation Scheme}\label{c_npe_s_npe_implementation_scheme}
To explore the implementation of NPE on ERGM, we have three implementation aims. Firstly, we want to provide an overall assessment of the appropriateness of adopting NPE for ERGM. To achieve that, we assess the general bias of the NPE and quantify the magnitudes of those biases. Secondly, we are interested in exploring and highlighting situations where NPE may perform particularly challenging on ERGM. This means that for those implementations that we found to be subject to large biases, we investigate the root causes of them and generalise them into scenarios that should be treated with caution. Finally, we want to assess and compare the NPE performance with the standard Bayesian ERGM.

For the implementation of SNPE, we want to explore how SNPE use iterative refining to better target a posterior for an ERGM. Furthermore, we assess the coverage issue to investigate the SNPE model behaviour when our initial prior or proposal does not provide an adequate coverage of the \enquote{true} posterior. While in other applications, this may lead to leakage issues  \citep{tejero2020sbi,deistler2022truncated}, we are interested in exploring how this may impact the ERGM implementations.

To achieve these research interests, we separately outline our implementation setups accordingly in this section.

\subsubsection{Main Bias Evaluation}
We start by outlining the main bias evaluation scheme of NPE, which offers a general assessment on the overall appropriateness of implementing NPE on ERGM, and it also helps to highlight when NPE may perform poorly on ERGM.

We adopt a synthetic data analysis where we know the \enquote{true} parameter. However, conducting a fair evaluation of bias is non-trivial. We justify our considerations after outlining the evaluation approach.

To assess the bias of the fitted NPE, we consider a set of $K$ \enquote{true} parameters, denoted as $\theta_{\text{true}}^k$ with $k = 1, \dots, K $. For each $\theta_{true}^{k}$, we simulate $M$ data (summary statistics), denoted as $x_{m}^{k}$ for $m=1,...,M$.

We fit NPE on every $x_{m}^{k}$ and take its posterior mean $\hat{\theta}_{m}^{k}$. We then take the average across all $M$ estimated posterior means to yield the final point estimate to enable bias assessment:
$$\hat{\theta}^{k}=\dfrac{1}{M}\sum_{m=1}^{M}\hat{\theta}_{m}^{k}.$$
We now justify the bias evaluation scheme and the selection of $M$ and $K$. Based on the overall computation requirements, we consider an overall of 40 $\theta_{true}$ (i.e., $K=40$).

To ensure the results generalisability and fairness of the evaluation, we select the 40 evaluation cases across various network topologies in our targeting graph space. More specifically, we are only interested in the sparse graph space with graphs with less than 1,100 edges (for a 90-vertex network), which is justified and defined from an ERGM inference perspective in the Supplementary Materials. We randomly sample parameters that produce networks with edge counts in four strata (in each stratum, we sample 10 parameter sets): $[0,275)$, $[275,550)$, $[550,825)$, $[825,1100)$. Since a variety of network characteristics depend on the network density (e.g., denser networks often show higher values in some statistics \citep{impactdensi}), our synthetic data simulation scheme gives an adequate network configuration coverage for generalisable findings.

We now justify the evaluation approaches. A single realisation of $x_{obs}$ may not be representative of its corresponding $\theta_{true}$, which is subject to the ERGM density variances. Especially, the mapping (through ERGM density) between parameters and realised statistics is neither one-to-one nor necessarily unimodal. Without averaging out the impact of ERGM variance, our evaluated bias is not specific to the NPE bias only. Therefore, we achieve the reliability by averaging across $M$ realisations, where the selection of $M$ is a balance between computational burden and effectiveness. We use the Wilcoxon Rank-Sum test to compare between distributions based on different simulation sample sizes. Based on this analysis, we set $M = 1,000$.

Overall, we implement $40,000$ fittings of NPE ($M=1,000$ and $K=40$). For each fitting, we draw $100,000$ posterior samples.

We use three metrics to measure and present the calculated biases:

1) Mean Error (ME):\\
$$\text{ME} = \frac{1}{K} \sum_{k=1}^{K} (\theta_{\text{true}}^k - \hat{\theta}^k).$$
2) Mean Absolute Error (MAE):\\
$$\text{MAE} = \frac{1}{K} \sum_{k=1}^{K} |\theta_{\text{true}}^k - \hat{\theta}^k|.$$
3) Root Mean Square Error (RMSE):\\
$$\text{RMSE} = \sqrt{\frac{1}{K} \sum_{k=1}^{K} (\theta_{\text{true}}^k - \hat{\theta}^k)^2}.$$
In the implementation of NPE, we use MAF as the conditional density estimator and consider 50 hidden units and 5 transformations (see Supplementary Material for selection consideration and recommendations). For the amortised NPE implementation (as well as the Bayesian ERGM), we consider a wide multivariate Normal prior distribution:
$$\pi(\theta) \sim \mathcal{N}\!\Bigl(
  \begin{pmatrix}
    0\\
    0\\
    0
  \end{pmatrix},
  \begin{pmatrix}
    10 & 0  & 0\\
    0  & 10 & 0\\
    0  & 0  & 10
  \end{pmatrix}
\Bigr).$$
The selection of the wide non-informative prior is justified by the diverse synthetic analysis cases, and to further incorporate such diversity, we select the training data-parameter pairs as $B=500,000$. The proposal distribution is chosen to be the same as the prior distribution for amortised NPE.

\subsubsection{Magnitude of Bias}
Our next study aim is to quantify the magnitudes of the assessed NPE biases. However, the absolute discrepancies in the ERGM parameter space do not offer much interpretable information on the magnitude of NPE bias. We therefore map back to the graph space (summary statistics space) to quantify whether the bias is significant. In particular, we want to provide a binary definition of whether a fitted NPE provides a \enquote{good} or \enquote{bad} estimate.

More specifically, for each case of $\theta_{\text{true}}^{k}$, we generate data (network samples) representing the \enquote{true} data distribution,
$$\mathbf{x}^{k}_{\text{true}} = \{ x_{\text{true}, 1}^{k}, \dots, x_{\text{true}, 100000}^{k} \}.$$
This is used to compare with the \enquote{posterior predictive} results from the NPE fittings, where we realise data from the posterior mean of each estimated $\hat{\theta}_{m}^{k}$ for $m=1,...,1000$ to account for the inherent ERGM model variance. We generate $100$ realisations from each $\hat{\theta}_{m}^{k}$, giving an overall $100,000$ realisations,
$$\mathbf{x}^{k} = \bigcup_{m=1}^{1000} \{ x_{m, 1}^{k}, \dots, x_{m, 100}^{k} \}.$$
We compare $\mathbf{x}^{k}$ to $\mathbf{x}^{k}_{\text{true}}$. Our main interest is in the estimation coverage instead of the exact shape of the distributions. To quantify the magnitude of bias in the data space, we define the bias as small if the mean of $\mathbf{x}^{k}$ lies within the $5\%$ and $95\%$ quantiles of $\mathbf{x}^{k}_{\text{true}}$, applying to each statistic. We also assess the percentage of coverage for cases classified as having large bias.\\

\subsubsection{Compare with Bayesian ERGM}
We are also interested in comparing the NPE fittings with the standard Bayesian ERGMs, which is the target model our fitted NPE aims to match.

We use the package \texttt{Bergm (5.0.7)} to fit Bayesian ERGMs. We set a burn-in period of 1,000 iterations for each fit and generate 6,000 posterior samples. The Bayesian ERGM uses the same prior distribution as the amortised NPE for consistency.

Similar to NPE, we account for the inherent ERGM variance by repetitively fitting the Bayesian ERGM on 1,000 realisations, and we compute and compare the posterior means of each fitting.

The computational demands of fitting Bayesian ERGM are significantly higher than those of NPE due to the lack of amortisation. This makes it infeasible to explore all 40 $\theta_{\text{true}}$ values or to increase the number of iterations and posterior samples. Therefore, we consider four $\theta_{\text{true}}$ cases for comparison (resulting in a total of 4,000 Bayesian ERGM fits). We choose two $\theta_{\text{true}}$ cases where NPE shows small to moderate bias and two cases where NPE exhibits significantly large bias (Table~\ref{tab:c_npe_bergm_four_cases}). In particular, Case 3 corresponds to the $\theta_{\text{true}}$ value with the highest NPE bias among all 40 tested cases.

\begin{table}[H]
\centering
\small
\begin{tabular}{lccc|ccc}
\toprule
 & \multicolumn{3}{c}{\textbf{Case 1}} & \multicolumn{3}{c}{\textbf{Case 2}} \\
\cmidrule(r){2-4} \cmidrule(l){5-7}
 & Edges & GWESP & GWNSP & Edges & GWESP & GWNSP \\
\midrule
\(\theta_{\text{true}}\) & 0.37  & -0.28 & 6.79  & -1.54 & -1.01 & 0.72 \\
Absolute Error                     & 0.20  & 0.03  & 0.09  & 0.03  & 0.01  & 0.01 \\
\bottomrule
\end{tabular}

\vspace{1em} 

\begin{tabular}{lccc|ccc}
\toprule
 & \multicolumn{3}{c}{\textbf{Case 3}} & \multicolumn{3}{c}{\textbf{Case 4}} \\
\cmidrule(r){2-4} \cmidrule(l){5-7}
 & Edges & GWESP & GWNSP & Edges & GWESP & GWNSP \\
\midrule
\(\theta_{\text{true}}\) & -11.96 & 0.27 & 1.05  & -5.79 & 1.13  & 0.61 \\
Absolute Error                     & 5.19   & 0.15 & 0.41  & 1.01  & 0.34  & 0.09 \\
\bottomrule
\end{tabular}
\caption{Selected $\theta_{\text{true}}$ values and evaluated NPE bias using absolute error. Each number corresponds, in order, to the statistics: number of edges, GWESP, and GWNSP; four cases are considered; Case 1 and Case 2 are for small to moderate NPE biases; Case 3 and Case 4 are the very high NPE bias cases.}
\label{tab:c_npe_bergm_four_cases}
\end{table}

\subsection{SNPE Implementation}
For the SNPE, we want to explore how to use iterative refining to better target a posterior. Furthermore, we assess the importance of coverage.

\begin{center}
\begin{figure}[H]
\centering
\includegraphics[width=4in, height=4.3in]{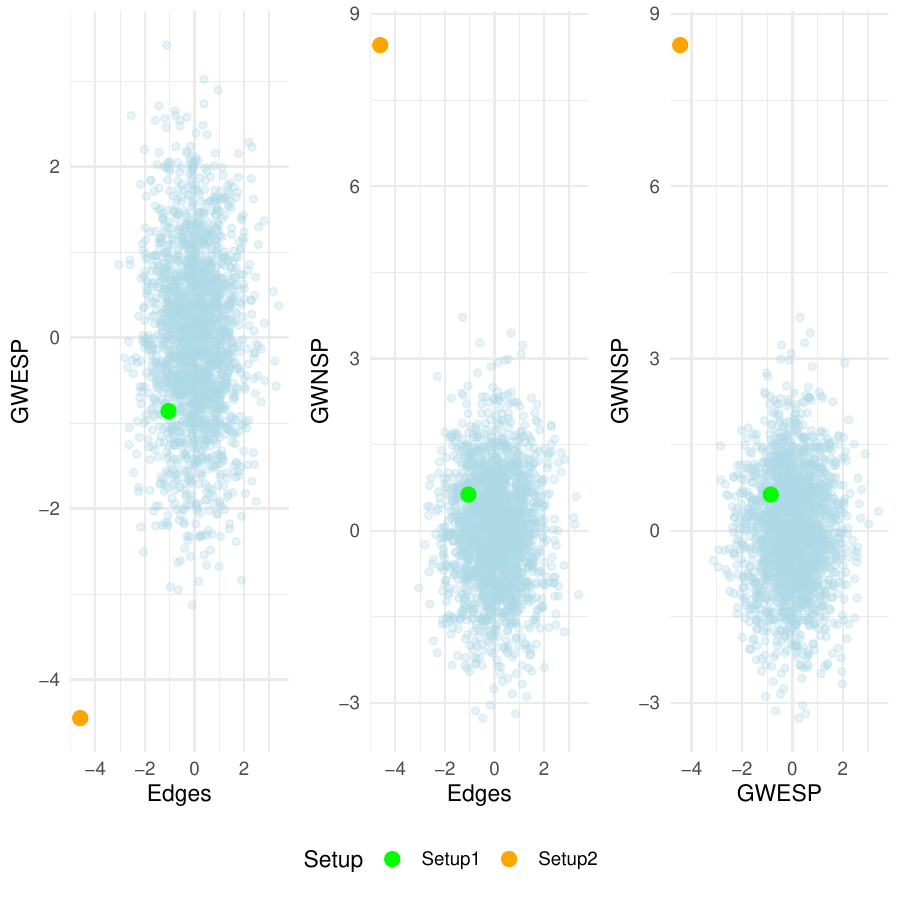}
\caption{Illustration of selected $\theta_{\text{true}}$ cases to implement SNPE; Case 1 is the green point, and the parameters are: -1.05, -0.86, and 0.63; Case 2 is the orange point, and the parameters are: -4.61, -4.45, and 8.46; the domain of blue points represents the coverage provided by the prior.}
\label{fig:c_npe_s_snpe_setup}
\end{figure}
\end{center}

More specifically, we perform a similar simulation study and consider only two cases of true parameters $\theta_{true}$, which are presented in Figure~\ref{fig:c_npe_s_snpe_setup} (Setup 1 and Setup 2). To justify, SNPE loses the amortisation property and we can only infer on particular observations. To assess coverage, we consider a more constrained prior distribution (and thus the initial proposal distribution), using $\pi(\theta) \sim \mathcal{N}(0,I)$.

In Setup 1, we randomly choose a point within the coverage of the prior distribution, aiming to explore and assess how SNPE iteratively refines modelling. We consider two implementations in Setup 1. One implementation uses more data-parameter pairs for training with $B=100,000$. We perform 5 SNPE rounds, constituting an overall $500,000$ data-parameter pairs for the conditional density estimator training.

To better assess SNPE behaviour when the initial estimations are crude, it is compared with another implementation using a small $B=1,000$, which is referred to as the Setup 1 reduced. We present 8 rounds of SNPE on Setup 1 reduced.

In Setup 2, we randomly select a \enquote{true} parameter outside the coverage of the prior distribution, aiming to assess SNPE behaviour when the coverage is not appropriate. The implementation scheme is the same as the Setup 1.\\
For the SNPE, we use the same conditional density estimator MAF with the same architecture as the amortised NPE.

\section{Implementation Result}\label{c_npe_s_npe_result}

\subsubsection{Main Bias Evaluation}
We fit the amortised NPE as described in the implementation scheme, and the evaluated biases are presented in Table~\ref{tab:overall_bias_table} (A) using three metrics.

Although direct interpretation of absolute biases in the parameter spaces can be abstract, we can conclude the assessed NPE biases are generally small, and such a conclusion is reliable and NPE-specific as we have averaged out the ERGM variances.

Furthermore, we observe that ME and MAE are quite different in scales. This means the NPE biases tend to be symmetric as symmetric biases may be cancelled out in ME.

The RMSE penalises more heavily for more extreme errors, and we observe that the bias is larger in RMSE. While most assessed cases are subject to small NPE biases, there exist cases with extreme biases. More specifically, we find 12 cases of $\theta_{\text{true}}$ where the absolute magnitude of NPE bias is larger than 1.5 for at least one of the three parameters (of summary statistics). We find that 7 out of these 12 cases exhibit multi-modal behaviours.

Moreover, our analysis finds that NPE performs relatively poorly for cases in the density stratum $[0,275)$. 7 out of 10 cases in that stratum characterise graph spaces located near the empty graph. This highlights the challenges of what we refer to as \enquote{boundary effects}. We will give more in-depth discussions of those challenges in Section~\ref{c_npe_s_npe_discussion}.

\subsubsection{Magnitude of Bias}
The previously discussed biases in the parameter space are abstract, and therefore, we project the NPE results to the data space (graph space) to yield more interpretable bias measurements. This mapping may be perceived as assessing posterior predictive.

Based on our designed significance criteria (Section~\ref{c_npe_s_npe_implementation_scheme}), we can see the majority of cases are accurate under NPE (Table~\ref{tab:overall_bias_table} (B)). Interestingly, the bias in GWESP is more severe than in the other two statistics. A possible explanation is that GWESP maybe more sensitive in the sparse domain, but we have not rigorously validated the actual cause.

We then focus on the cases regarded as \enquote{high} or \enquote{significant} biases reflected in the data space. In Table~\ref{tab:overall_bias_table} (C), we assess the predictive coverage percentages between $\mathbf{x}^{k}_{\text{true}}$ and $\mathbf{x}^{k}$ for those cases. We observe that NPE still provides reasonable predictive value coverage for most cases, typically with one statistic showing relatively worse coverage, meaning those cases are not completely inaccurate. More importantly, in five out of eight large-bias cases, the \enquote{true} observed statistics lie near the extreme end of the statistics space (typically near zero). This again highlights the \enquote{boundary effect}.

Overall, when quantifying bias in both the parameter space and graph space, we find that the NPE performs well in most cases.

\begin{table}[H]
\centering
\small
\begin{tabular}{lccc}
\toprule
\multicolumn{4}{l}{\textbf{A. Bias Metrics}} \\
\midrule
& \textbf{Edges} & \textbf{GWESP} & \textbf{GWNSP} \\
\cmidrule(lr){1-4}
ME   & -0.01 & -0.30 & -0.09 \\
MAE  & 0.61  & 0.61  & 0.38 \\
RMSE & 1.09  & 1.04  & 0.82 \\
\addlinespace
\midrule
\multicolumn{4}{l}{\textbf{B. Insignificant Bias Percentages}} \\
\cmidrule(lr){1-4}
\makecell[l]{Insignificant\\ bias rate} & $92.5\%$ & $82.5\%$ & $95.0\%$ \\
\addlinespace
\midrule
\multicolumn{4}{l}{\textbf{C. Significant Bias Cases Predictive Coverage}} \\
\cmidrule(lr){1-4}
\textbf{Case (k)} & \textbf{Edges} & \textbf{GWESP} & \textbf{GWNSP} \\
\midrule
\textbf{2}  & $60.31\%$ & $0.02\%$ & $71.93\%$ \\
\textbf{5}  & $79.36\%$ & $0.68\%$ & $81.43\%$ \\
\textbf{8}  & $3.71\%$ & $65.33\%$ & $100.00\%$ \\
\textbf{9}  & $45.67\%$ & $38.17\%$ & $47.92\%$ \\
\textbf{10} & $34.58\%$ & $29.58\%$ & $36.32\%$ \\
\textbf{11} & $58.79\%$ & $31.47\%$ & $64.64\%$ \\
\textbf{13} & $39.57\%$ & $39.88\%$ & $23.55\%$ \\
\textbf{20} & $84.42\%$ & $13.37\%$ & $79.65\%$ \\
\bottomrule
\end{tabular}
\caption{Table summarising the evaluated NPE biases and their bias magnitudes quantification; (A)~Evaluated NPE biases summarised using three metrics for all 40 cases: ME, MAE, and RMSE;
(B)~The percentages of cases that are regarded as insignificant bias using our quantification criteria in the data space;
(C)~The percentages of summary statistics density coverage for cases regarded as significant or large biases.}
\label{tab:overall_bias_table}
\end{table}

\subsubsection{Compare with Bayesian ERGM}
Finally, we want to compare our NPE fittings with the Bayesian ERGM, and as explained and justified in Section~\ref{c_npe_s_npe_implementation_scheme}, we consider four cases as outlined in Table~\ref{tab:c_npe_bergm_four_cases}.

\begin{center}
\begin{figure}[H]
\centering
\includegraphics[width=4.5in, height=4.5in]{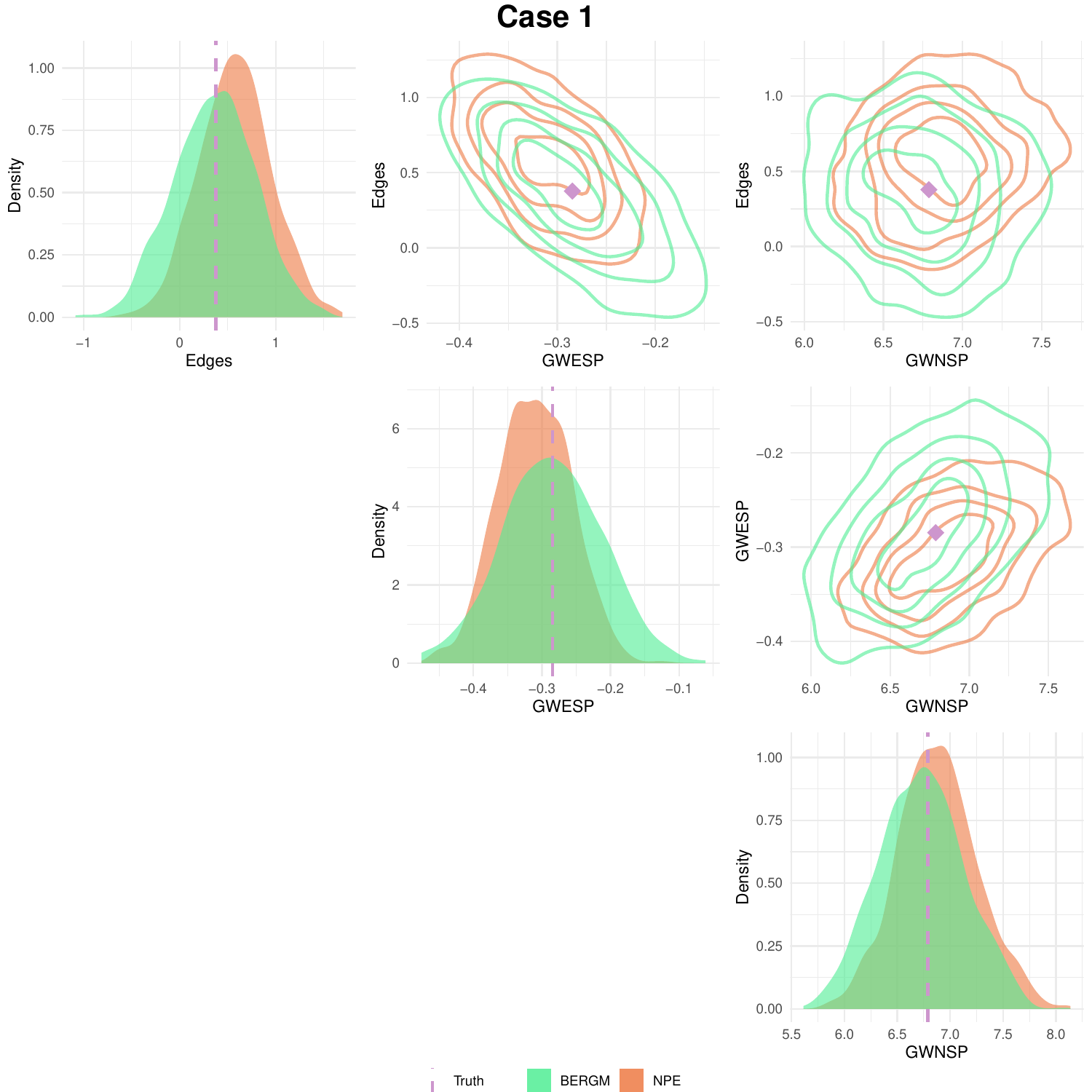}
\includegraphics[width=4.5in, height=4.5in]{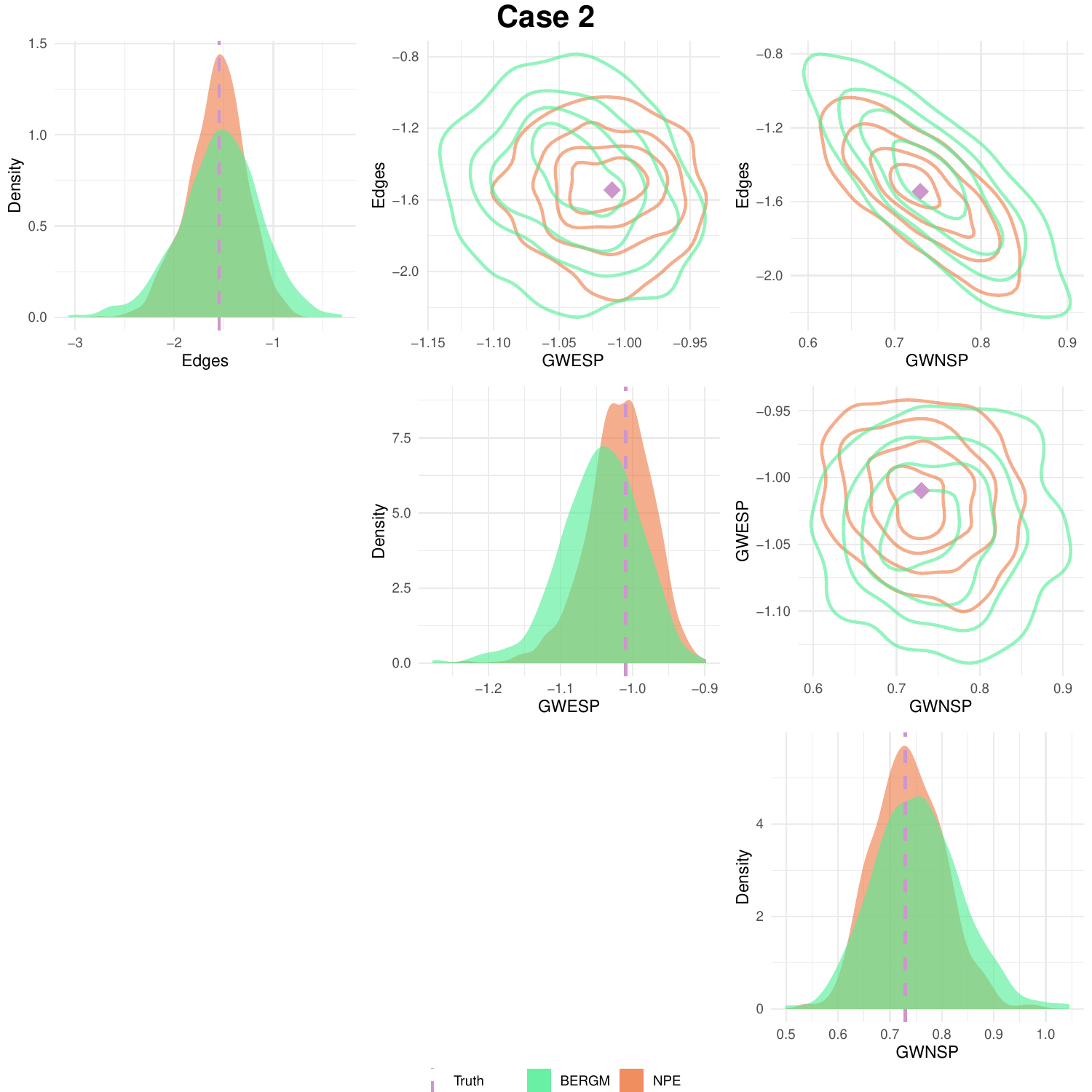}
\caption{Plots for comparing between NPE and Bayesian ERGM Case 1 and Case 2; posterior density plots for parameters edges, GWESP and GWNSP (diagonal); Contour plots for pairwise posterior densities spaces (off-diagonal); plotted for posterior means for NPE (orange) and Bayesian ERGM (green); the corresponding \enquote{truth} is plotted as purple (dashed lines and points).}
\label{fig:c_npe_s_npe_compare_bergm_case12}
\end{figure}
\end{center}

In Cases 1 and 2 (Figure~\ref{fig:c_npe_s_npe_compare_bergm_case12}), the fitted NPEs and Bayesian ERGMs span similar posterior parameter spaces, centred around the \enquote{true} parameter values. In particular, the fitted NPEs successfully capture the correlation tendencies between parameters (e.g., between edges and GWNSP in Case 2), mirroring the behaviour of Bayesian ERGMs. Given these results, we find NPE performs as well as the Bayesian ERGM.

\begin{center}
\begin{figure}[H]
\centering
\includegraphics[width=4.5in, height=4.5in]{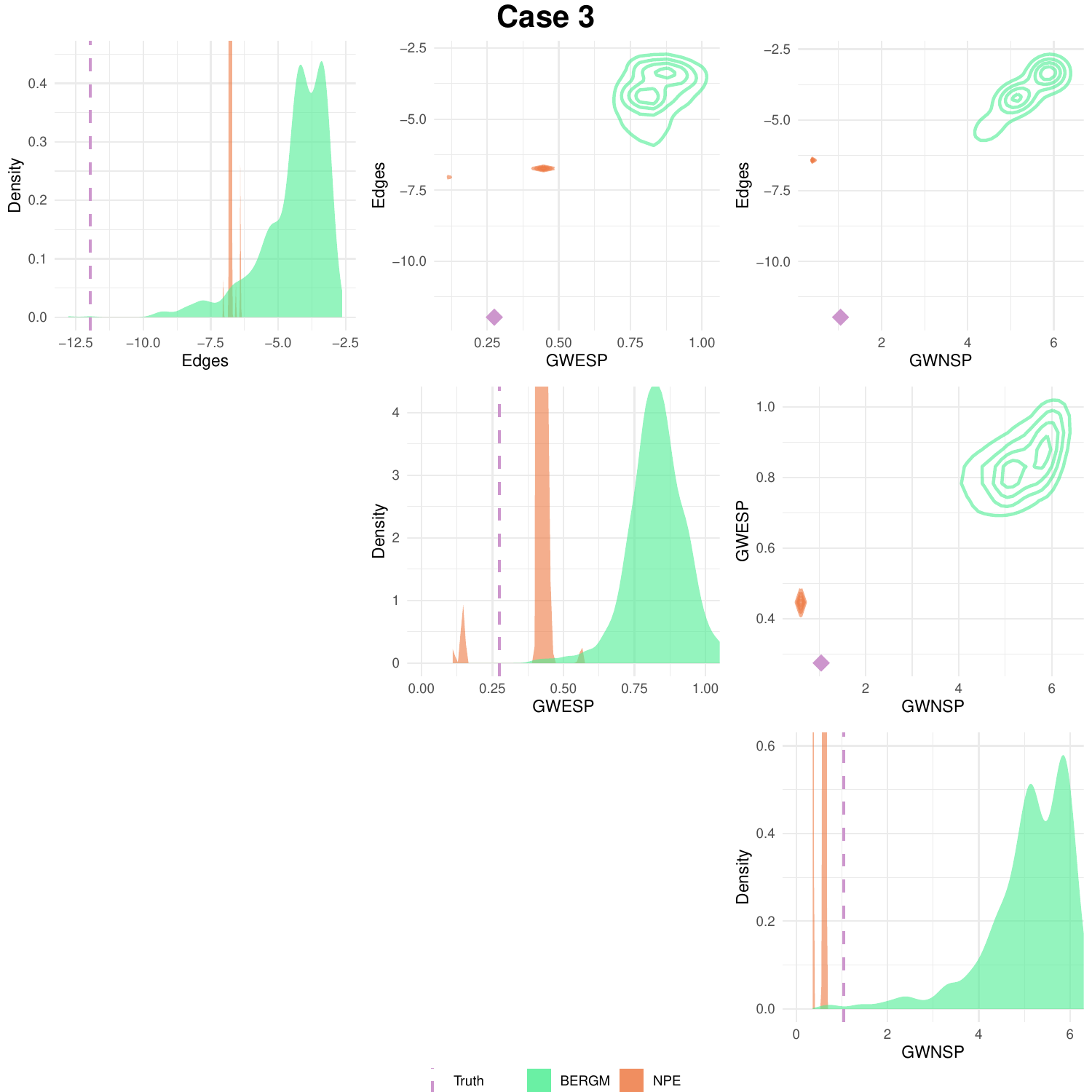}
\caption{Plots for comparing between NPE and Bayesian ERGM Case 3; posterior density plots for parameters edges, GWESP and GWNSP (diagonal); Contour plots for pairwise posterior densities spaces (off-diagonal); plotted for posterior means for NPE (orange) and Bayesian ERGM (green); the corresponding \enquote{truth} is plotted as purple (dashed lines and points).}
\label{fig:c_npe_s_npe_compare_bergm_case3}
\end{figure}
\end{center}

Case 3 shows a situation in which the \enquote{true} parameter set yields extreme statistics near the empty network, making NPE subject to the largest bias among all 40 cases of $\theta_{true}$. As shown in Figure~\ref{fig:c_npe_s_npe_compare_bergm_case3}, Bayesian ERGMs also perform poorly in estimating $\theta_{true}$, showing significant deviations from the \enquote{true} parameter. This case showcases the challenges of the \enquote{boundary effect}, where a large parameter space may correspond to the near-extreme graphs, making model training and inference difficult.

\begin{center}
\begin{figure}[H]
\centering
\includegraphics[width=4.5in, height=4.5in]{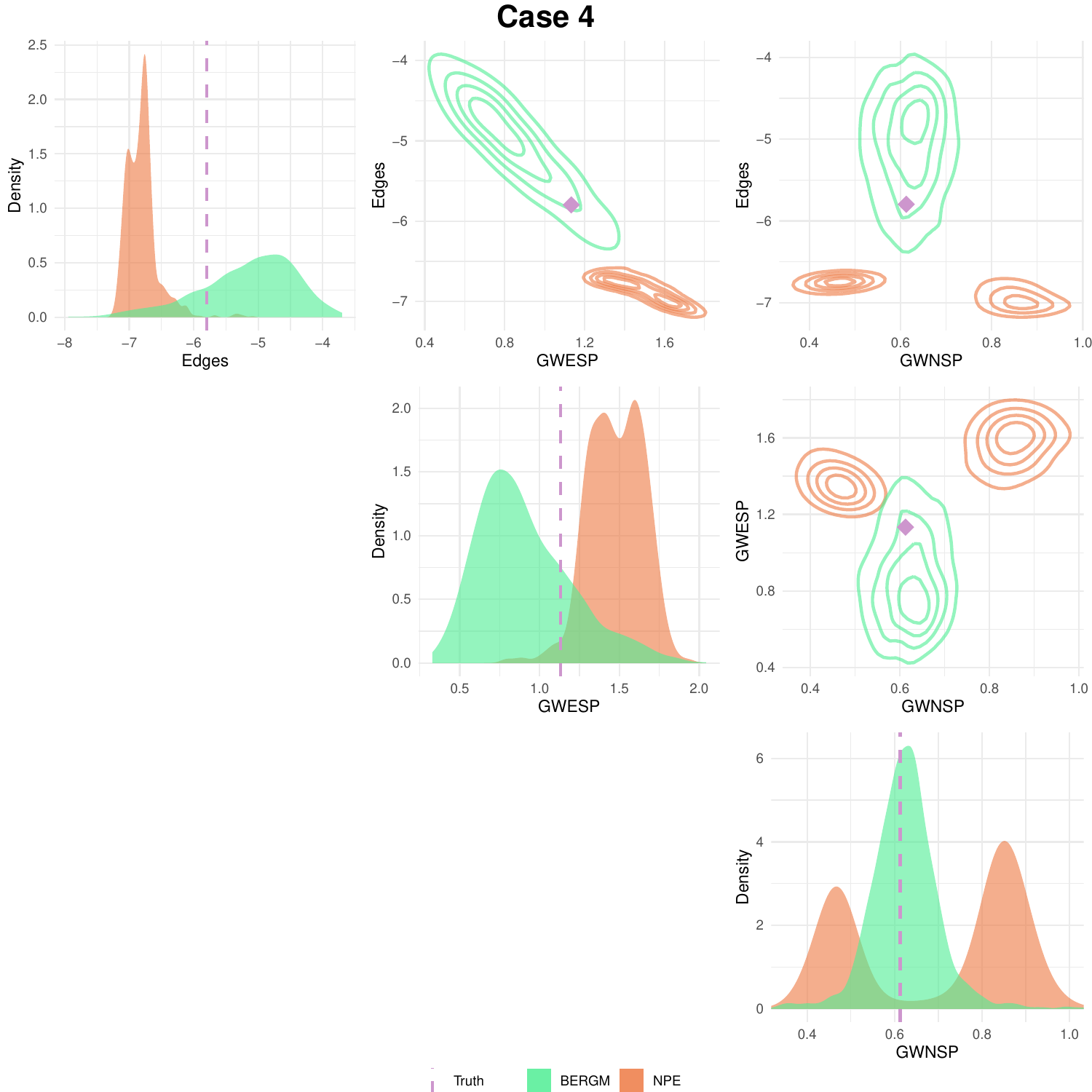}
\caption{Plots for comparing between NPE and Bayesian ERGM Case 4; posterior density plots for parameters edges, GWESP and GWNSP (diagonal); Contour plots for pairwise posterior densities spaces (off-diagonal); plotted for posterior means for NPE (orange) and Bayesian ERGM (green); the corresponding \enquote{truth} is plotted as purple (dashed lines and points); \textbf{Explanation}: Case 4 is a bimodal case, especially in GWNSP. Both the NPE and Bayesian ERGM perform poorly. Case 4 shows bi-modality when initialising at the full network and using 50,000 network simulation iterations. Adjusting the simulation setup may change the distribution, which will be discussed later.}
\label{fig:c_npe_s_npe_compare_bergm_case4}
\end{figure}
\end{center}

In Case 4 (Figure~\ref{fig:c_npe_s_npe_compare_bergm_case4}), we observe bi-modality in the data space corresponding to this $\theta_{true}$. Namely, the bi-modality mainly shows in GWNSP, where we can find that NPE shows two separate clusters of parameter estimates. Both the Bayesian ERGM and NPE show poor performance on the estimation, with posterior (mean) densities showing significant deviation from the \enquote{true} parameter.

Overall, we have shown that NPE can achieve the same level of accuracy as the Bayesian ERGM. However, we should be more cautious in cases that are naturally difficult to estimate.

\subsection{SNPR Implementation Result}

\subsubsection{Setup 1}
\begin{center}
\begin{figure}[H]
\centering
\includegraphics[width=4in, height=5in]{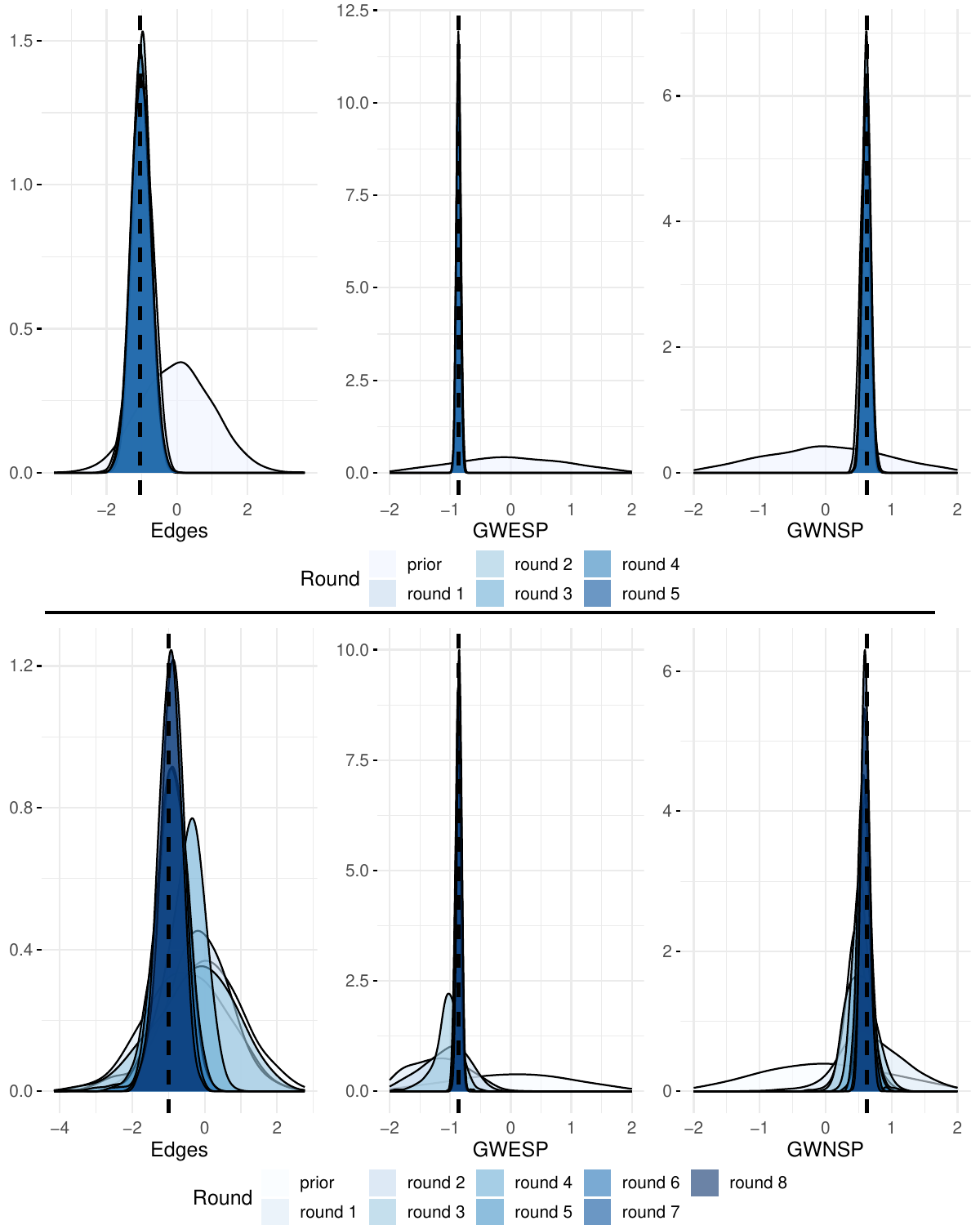}
\caption{Density plots for posterior densities across SNPE rounds; later rounds are presented with less transparencies and deeper colours; the plots are for the SNPE Setup 1 (above) and Setup 1 reduced (below); the Setup 1 shows 5 rounds; the Setup 1 reduced shows 8 rounds; each density plot considers 100,000 posterior samples; the \enquote{true} parameters are plotted as the black dashed lines.}
\label{fig:c_npe_snpe_implementation_case1_results}
\end{figure}
\end{center}

In Setup 1, the target observation corresponds to \enquote{true} parameter values that locate within the coverage of the prior distribution (Figure~\ref{fig:c_npe_s_snpe_setup}). In Figure~\ref{fig:c_npe_snpe_implementation_case1_results} (above), where we use $B=100,000$, the posterior densities stabilise after round 1 (round 1 is equivalent as fitting amortised NPE), with no further refinement in later rounds. This demonstrates a case where the samples (from the prior distribution) used to train the first-round SNPE are sufficient to give an accurate estimation, indicating that a large $B$ is unnecessary, especially when the coverage is now narrower (than our amortised NPE implementations).

We therefore, challenge a further case such that the sequential training size is small. In Figure~\ref{fig:c_npe_snpe_implementation_case1_results} (below), we observe a track of estimation improvement across SNPE rounds. In particular, the variance of the posterior density is increasingly refined and decreased across rounds. SNPE can successfully explore the parameter space to yield an adequate posterior estimation with iterative refinements. SNPE is also a more efficient approach and requires a smaller overall number of simulations needed to achieve accurate estimations \citep{greenberg2019automatic}. In Setup 1 reduced, only a total of $8,000$ iterations is used to reach convergence.

\subsubsection{Setup 2}
\begin{center}
\begin{figure}[H]
\centering
\includegraphics[width=4in, height=5in]{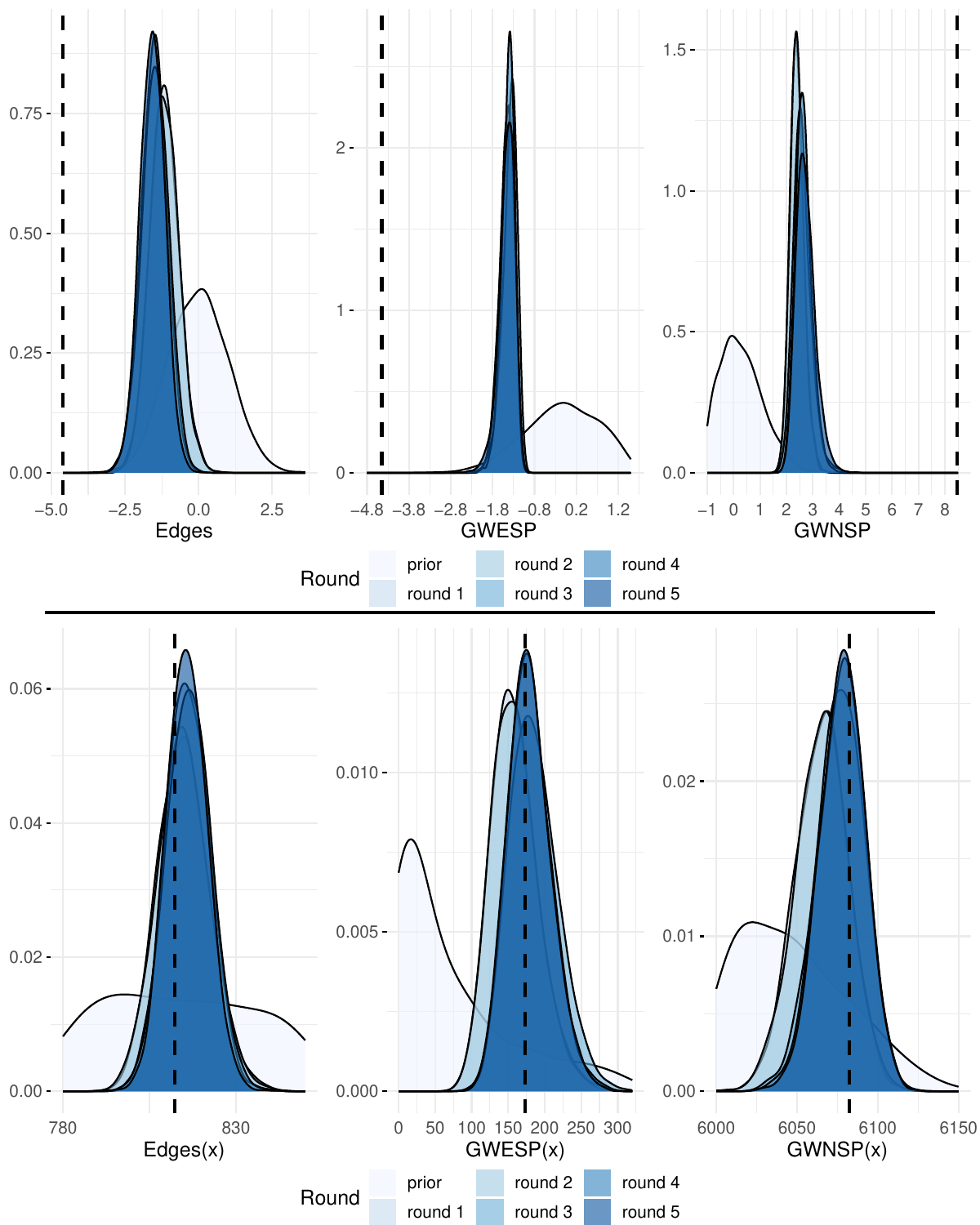}
\caption{Density plots for posterior densities across 5 SNPE rounds for the Setup 2 (above); density plots for posterior predictive across 5 SNPE rounds (below); later rounds are presented with less transparencies and deeper colours; each density plot considers 100,000 posterior samples and 5,000 posterior predictive samples; the \enquote{true} parameters and observed statistics are plotted as the black dashed lines.}
\label{fig:c_npe_snpe_implementation_case2_results}
\end{figure}
\end{center}

In Setup 2, we consider a case where the coverage is not sufficiently wide to provide support for the \enquote{true} posterior. In Figure~\ref{fig:c_npe_snpe_implementation_case2_results} (above), we can observe that SNPE cannot accurately estimate the \enquote{true} parameter, although the SNPE estimates converge in the \enquote{correct} directions of summary statistics tendencies. This is because SNPE uses atomic loss to return a distribution that is only proportional to the true posterior in support of the proposal.

However, in other applications, insufficient coverage may lead to leakage issues \citep{tejero2020sbi,deistler2022truncated}. Through our attempts, our implementation of SNPEs on ERGMs did not encounter severe leakage issues when the support was not sufficiently wide. We assess the posterior predictive of the biased SNPE estimates to explore a possible reason. We find the posterior predictive resides relatively close to the observed value (Figure~\ref{fig:c_npe_snpe_implementation_case2_results} below). This is because the inter-relation between summary statistics \enquote{convinces} the SNPE to use an \enquote{alternative}, which can still yield similar (but not exact) posterior predictive. This highlights an issue when the coverage is not sufficiently wide. In ERGM implementation, without knowing the \enquote{true} parameter, the coverage issue may be masked by the seemingly adequate posterior predictive. The coverage issue also has other implications on ERGM implementations, which we will discuss later.

\section{Discussion}\label{c_npe_s_npe_discussion}
Through our implementations of NPE and SNPE on ERGMs, we have established their appropriateness. We have shown that the estimations can achieve adequate accuracies when targeting a variety of network topological structures. Both evaluations of NPE biases in the parameter space and the graph space demonstrate accuracies. More importantly, using NPE can yield equivalently good performance as standard Bayesian ERGMs. This fulfills our main research aim of assessing the use of NPE on ERGMs.

Furthermore, our study has demonstrated the core advantage of using an amortised NPE on ERGMs, which is the scalability. In Section~\ref{c_npe_s_npe_implementation_scheme}, we have outlined a bias evaluation scheme with an overall $40,000$ fittings, each drawing $100,000$ posterior samples, which itself consists of four-billion computationally intensive exchanging steps if using standard Bayesian ERGM (see Algorithm~\ref{alg:approximate_exchange}). Such scalability is due to the amortisation and the invertibility of the trained conditional density estimators (see Supplementary Material), and using standard Bayesian ERGM estimation is infeasible to achieve such large-scale implementations. This has significant implications for the recent advancements in ERGM, where we need to infer on a population of subjects instead of only one \citep{Lehmann665398}.

While NPE and SNPE can perform well on most ERGM cases, we highlight several issues that can lead to challenges upon implementation, which are multi-modality, \enquote{boundary effects} and coverage issues. They form the ERGM-specific challenges and such a summary contributes to future studies to diagnose their NPE fitting difficulties on ERGMs.\\

\subsubsection{Multi-modality}
In ERGM context, multi-modality refers to situations where the same set of parameters can equivalently characterise (through ERGM density) two or more distinct areas of the graph space. In practice, the multi-modality issue impacts the quality of training data, and the data (network) parameter pair realisations become sensitive to the realisation procedure when realising for multi-modal parameters.

When realising a network, we are essentially exploring the graph space to search for a domain in the graph space that matches the ERGM parameters (see Supplementary Material). When subject to multi-modality, we often see sharp \enquote{jumps} in the graph space, and in particular, we often observe the exploration process \enquote{trapped} in dense network modes rather than in the sparse network modes. Therefore, the realisation process is sensitive to the network realisation iterations and initialisations. More specifically, when we initialise the graph space exploration at a dense network, we tend to be \enquote{trapped} in the first dense network mode corresponding to the parameter, without exploring the sparse network modes. In contrast, when we initialise at a sparse network, we can initially explore the sparse network modes, and if the realisation iteration is large enough, it will again be \enquote{trapped} in the dense network modes. Furthermore, the multi-modality issue is also dependent on the choice of summary statistics combinations.

This has two implications for implementing NPE on ERGM. First, the NPE inference quality and accuracy can be sensitive to the choice of realisation initialisation and iterations. We need to ensure consistency in these aspects when replacing Bayesian ERGM with NPE. Secondly, the quality of the training dataset is compromised when containing a lot of multi-modal parameters, and this multi-modality issue tends to be more severe when targeting a very sparse graph space. The density estimators may be \enquote{confused} by these data-parameter pairs and may show poor performance or suffer from posterior leakage, especially when using the atomic loss in SNPE. We may enhance the expressiveness of the density estimator or adopt a deeper flow design to target a training set of poor quality.

\subsubsection{Boundary Effect}
Another issue we highlight is the \enquote{boundary effect}. We use \enquote{boundary effect} to refer to the large parameter space mapping to the extreme network configurations (i.e., full or empty networks). In particular, once the ERGM parameter set \enquote{hits} the boundary of extreme networks, any further increase in parameter magnitude will continue to generate the same extreme networks. For example, if the number of edges parameter equals 1 can yield fully connected networks; further exploration of the parameter space in this direction (e.g., setting it to 2) will still yield fully connected networks. This issue becomes very sensitive and complex under the incorporation of multiple summary statistics. This is related to the famous degeneracy issue in ERGM \citep{mele2017structural,degeeeen2,degeen4,degeen3}, which describes an unstable parameter space such that a minor increase in some parameters can rapidly and unstably push the ERGM density towards the extreme graph space.

The \enquote{boundary effect} often leads to sharp transitions or plateaus in the probabilistic mapping between ERGM parameters and data. In practice, we find using an expressive density estimator may be too rigid to model such behaviour. In particular, such expressive density estimators may over-fit these plateaus, representing them as wide, flat regions in the parameter space, and this may mask other useful relations. Through our attempts, we are more likely to suffer from leakage issues using more expressive density estimators when the proposal coverage is wide (because it covers a lot of parameter space with boundary effects and also multi-modalities).

\subsection{Coverage}
Through our implementations, we find the coverage issue impacts our ERGM estimation in two ways. Firstly, a biased estimate due to narrow coverage can still produce posterior predictions similar to the observed data. This is because the inter-relationships between ERGM summary statistics can yield similar summary statistics with different parameters, making SNPE less prone to leakage issues due to coverage but suffering from masking of \enquote{true} posterior exploration. Note that the alignment in posterior predictive achieved in such cases will not be \enquote{perfect}. Therefore, it is important to provide sufficiently wide coverage to avoid biased estimations.

Secondly, although we desire adequate coverage, an excessively wide coverage has a mixing impact with the \enquote{boundary effect} and the multi-modalities, which may make our NPE and SNPE implementations encounter other challenges.

\subsection{Limitations and Future Works}
While we have carefully justified each implementation procedure in this study, we discuss some inevitable limitations and how future studies may develop further.

Firstly, we have to highlight that using NPE is like throwing data-parameter sets into a \enquote{black box}. It is challenging to investigate the \enquote{true} root causes of the implementation difficulties and the leakage issues that we encountered during our explorations. We approach by auditing data inputs, monitoring loss functions and validating varying inference setups. Our summarised ERGM-specific issues rely on our attempts, experiences and understandings of ERGMs. We do not claim our summarised issues to be the absolute \enquote{truth}, but they are certainly useful for future studies. Future implementation of NPE on ERGM may focus on diagnosing their implementation difficulties according to our summarised issues.

While our study has pointed out some potential resolutions to the above issues, we have not systematically established implementation recommendations and guidelines in this study, which is not trivial. Some reinforcement learning-based NPE may be useful \citep{ward2022robust}. Future studies shall target these aspects accordingly.

In this study, we have not systematically explored the impact of conditional density estimators, neural network architecture and ERGM summary statistics on the performance of NPE. In practice, we have experimented by varying these features in our exploration, and provided general guidelines and our recommendations in the Supplementary Material, but these are beyond the scope and research interest of this study. Future studies may rigorously research in these directions further. Our study builds a solid foundation for future explorations.\\

\clearpage

\section{Supplimentary}
\subsection{Network Simulation Algorithm}
\begin{algorithm}[H]
  \caption{Network Simulation Algorithm With Given $\theta$}\label{alg:c_review_network_simulation}
  \begin{algorithmic}
    \State \textbf{Initialisation}
    \State Initialise a network $y^{(0)}$ of the same size as the targeting network\\
    \For {$i=1,...,I$}
    \State Propose $y'$ from $q(.|y^{(i-1)})$
    \State Accept $y'$ with\\
    \State $$\text{Acceptance Ratio}=\dfrac{p(y'|\theta)q(y'|y^{(i-1)})}{p(y^{(i-1)}|\theta)q(y^{(i-1)}|y')},$$\\
   
    \EndFor
    \State Note $q(.|y)$ can be as easy as a proposal of an addition or removal of an edge from $y^{(i)}$.
  \end{algorithmic}
\end{algorithm}

\subsection{Sufficiency proof}
Formally, we express the posterior distribution as:
$$\pi(\theta|y)=\dfrac{\pi(\theta)}{p(y)}\dfrac{exp(\theta h(y))}{\sum_{y'\in \mathcal{Y}} exp(\theta h(y'))},$$
and note that the whole expression only depends on $y$ through $h(y)$. More specifically, the prior \(\pi(\theta)\) is independent of $y$. The entire sampling distribution also depends only on $h(y)$. The marginal likelihood
$$p(y)
\;=\;
\int_{\Theta}
\pi(\theta)\,p(y \mid \theta)
\,d\theta,$$
is based on the prior and sampling distributions, and therefore, only depends on $h(y)$.\\

\subsection{Conditional Density Estimator} \label{c_npe_s_conditional_density_estimator}
Our implementation relies on the package \texttt{sbi (v0.22)} in python \citep{tejero2020sbi}, and we discuss some of the available conditional density estimators and establish some non-indepth guidelines on how to choose between them.\\

\textbf{Masked Autoregressive Flow (MAF)} is a normalising flow \citep{papamakarios2017masked}. We can think of it as we are \enquote{flowing} the probability through a set of transformations. In more details, it transforms a simple base distribution $z$ into our targeting complex distribution $x$ through an invertible and differentiable transformation function $f(.)$,
$$x = f(z),$$
$$p(x) = p(z) \left| \det \left( \frac{\partial f^{-1}(x)}{\partial x} \right) \right|,$$
where $\det \left( \frac{\partial f^{-1}(x)}{\partial x} \right)$ is the Jacobian determinant of the inverse transformation.\\
We now explain the other component of MAF. An autoregressive model allows factorisation of a multivariate joint distribution into a product of conditional distributions,
$$p(x_1, x_2, \dots, x_D) = \prod_{i=1}^{D} p(x_i \mid x_{1:i-1}).$$
This means each variable $x_i$ is predicted based on its previous variables.\\
We now combine them together to outline MAF. MAF uses an affine transformation,
$$x_{i} = z_{i} \exp(\alpha_{i}) + \mu_{i},$$
where
$$\mu_{i} = f_{\mu_{i}}(x_{1:i-1}),$$
$$\alpha_{i} = f_{\alpha_{i}}(x_{1:i-1}),$$
$$z_{i} \sim \mathcal{N}(0, 1).$$
$f_{\mu_{i}}$ and $f_{\alpha_{i}}$ are masked feedforward neural networks. They take inputs $x_{1:i-1}$, and output $\mu_{i}$ and $\alpha_{i}$.
$$p(x_{i} \mid x_{1:i-1}) = \mathcal{N}(\mu_{i}, \exp(\alpha_{i})^{2}).$$
We may adjust the architecture of the neural network to achieve certain properties. In our study, we only consider adjusting the number of hidden units to control the expressiveness of the flow.\\
Furthermore, we can add intermediate transforms of the same type to flow-based models so that the base distribution undergoes a series of flow transformations until it is transformed into the target distribution. We refer to this as the number of transformations, and it is good for modelling the complicated multi-modal behaviours in ERGM.\\

\textbf{Neural Spline Flow (NSF)} is also a normalising flow model. Instead of affine transformations, it uses spline-based transformations (monotonic rational quadratic splines) \citep{nsf}. It splits the input domain into intervals $[-B,B]$ using $K+1$ knots (think them as coordinate points):
$$\{u_0, \dots, u_K\},$$
and
$$\{v_0, \dots, v_K\}.$$
Now say our $x$ lands in the interval $[u_k, u_{k+1}]$, we can calculate the local knot slope,
$$ s^{(k)} = \frac{v_{k+1} - v_{k}}{u_{k+1} - u_{k}},$$
and then normalise it within the interval by
$$\xi = \frac{x - u_{k}}{u_{k+1} - u_{k}}.$$
Given the above, we can now define its transformation (in that interval) as 
$$f(x) = v_{k} + (v_{k+1} - v_{k}) \, \frac{s^{(k)}\xi^2 + \delta_{k}\,\xi(1-\xi)}{s^{(k)} + \Bigl(\delta_{k+1}+\delta_{k}-2s^{(k)}\Bigr)\,\xi(1-\xi)},$$
where \(\delta_{k}\) and \(\delta_{k+1}\) denote the derivatives at the knots. Neural networks are used to infer the knots and knot slopes. Similar to MAF, NSF can stack multiple transformation layers.\\

\textbf{Mixture Density Network (MDN)} considers a Normal mixture distributions with $P$ components,
$$q_{\phi}(\theta \mid x) = \sum_{i=1}^{P} w_{i}(x) \mathcal{N}\left(\theta \mid \mu_{i}(x), \sigma_{i}^{2}(x)\right).$$
$w_{i}(x)$ is the mixing weights summing to 1. $\mu_{i}$ and $\sigma_{i}$ are the mean and variance. A neural network (parametrised by \( \phi \)) is used to output $\{w_{i}(x), \mu_i(x), \sigma_i(x)\}_{i=1}^{P}$ using the input data $x$. The mixing weights $ w_i(x)$ use a softmax activation function to ensure summation to one, and the standard deviations $\sigma_i(x)$ use exponential or soft-plus activation functions. Put simply, MDN uses neural network to predict a mixture of Normal distributions to capture relationships between inputs and outputs.\\

In our study, we mainly use MAF, which we believe is sufficient for our study, and we tend to adopt deeper MAF flow (by increasing the number of transformations) to better capture complex ERGM densities (with sharp peaks and multi-modal distributions). Otherwise, we may adjust the number of hidden units to control the overall expressiveness and avoid over-fitting. As a general guideline, if (deep) MAF is still not expressive enough to model complex distributions, an NSF can be considered instead. The spline-based transformations in NSF can better capture complex non-linear relationships. In comparison, although MAF is also expressive, it is limited by the linearity of transformations. In addition, MDN may be used as well, and it is flexible to model multi-modal distribution using its mixture nature. MDN has a more important role in NPE, especially for SNPE, allowing a more tractable solution when the proposal distribution is chosen as Normal distribution-based densities.

\subsection{Atomic Proposal and Leakage Issue}
While SNPE correctly targets the posterior, the integral of the normalising constant
$$Z_{\phi}(x)=\int_{\theta} q_{\phi}(\theta|x)\dfrac{\tilde{p}(\theta)}{\pi(\theta)}$$
in Equation~\eqref{eq:c_npe_s_snpe_main_equation} is not always possible to evaluate, which depends on the choice of density estimator and proposal distribution. When the density estimator is MDN and the proposal is a Normal distribution-based distribution, SNPE can be used directly with a closed solution. Otherwise, SNPE uses an atomic proposal so that we can flexibly use different density estimators and proposal distributions. In particular, the flow-based density estimator used in this study requires atomic loss.\\
In atomic loss, we \enquote{replace} the integrals with sums. More specifically, we discretise the parameter space by considering finite sets (with mini-batch size $M$),
$$\Theta = \{\theta_1, \theta_2, \dots, \theta_M\},$$
$$\Theta \sim \tilde{p}(\Theta),$$
where the atomic set $\Theta$ is a finite subset of parameters sampled from \enquote{original} proposal (originally referred to as hyper-proposal in \citep{greenberg2019automatic}). We then define $U_\Theta(\theta)$ as a uniform distribution over $\Theta$.\\
Under the atomic set, the relationship between proposal posterior and actual posterior (Equation~\eqref{eq:c_npe_s_snpe_main_equation}) becomes
$$\tilde{q}_{\phi}(\theta|x) = \frac{q_{\phi}(\theta|x) / \pi(\theta)}{\sum_{\theta' \in \Theta} \left( q_{\phi}(\theta|x) / \pi(\theta') \right)}.$$
Under atomic loss, Greenberg et al. proved that using the expectation of a cross-entropy loss \citep{greenberg2019automatic},
$$L(\phi) = - \mathbb{E}_{\Theta, \theta, x} \left[ \log \tilde{q}_{x, \phi}(\theta) \right],$$
and then we can target the correct proposal posterior and the actual posterior. The expectation is evaluated based on $\Theta \sim \tilde{p}(\Theta)$, $\theta \sim U_\Theta(\theta)$ and $x \sim p(x|\theta)$.\\
In practice, $\Theta$ is sampled in mini-batches of size $M$ each representing an atomic set so that we can use standard gradient-based approaches for quick optimisations. The atomic set is produced from all simulated data-parameter pairs, and in SNPE, this includes those pairs from earlier rounds. See more details in \citep{greenberg2019automatic}.\\
With atomic loss, SNPE can be used flexibly with a range of density estimators, priors and proposals. Importantly, we address some requirements for the proposal distribution.\\ 
More importantly, we describe the leakage issue related to atomic loss. Posterior leakage refers to the situation where the estimated posterior distribution produced by a neural network assigns a lot of probability mass to regions of the parameter space that should have little probability under the true posterior distribution. A range of issues can lead to leakage. We discuss here in particular reasons related to the prior and proposal distributions. SNPE using atomic loss returns a distribution that is only proportional to the true posterior on the support of the proposal (or prior if starting with a prior). If the proposal is chosen excessively narrow (narrower than the prior), a leakage issue can happen \citep{deistler2022truncated}. Therefore, adequate coverage of posterior support is important.

\subsection{Sparse Graph Space Focus}
In our study, our synthetic data analysis focuses only on the sparse networks. This is because our chosen particular set of summary statistics exhibits strong collinearity when the networks are dense. Such collinearity is mainly due to the geometrically weighted features and is also encountered in past studies \citep{colinear1}. The consequence would be that the relationship between summary statistics tends to be more easily predicted, and this gives rise to a more certain relationship. Therefore, assessing the NPE in the dense range may cause model specification and identifiability issues, and such inference is not ideal.\\
We tested different cases of parameter combinations and estimated the sparse region to be less than 1,100 edges. Note that this size is only an approximation and applies to 90-vertex networks only using the summary statistics combination: edges, GWESP and GWNSP.

\printbibliography

\end{document}